# Comparative evaluation of catalyst materials using a binary choice model

Keiji Sakakibara[1,2] and Daniel M. Packwood[2*]

[1]Graduate School of Advanced Integrated Studies in Human Survivability (GSAIS), Kyoto University, Kyoto 606-8306, Japan

[2]Institute for Integrated Cell-Material Sciences (iCeMS), Kyoto University, Kyoto 606-8501, Japan

* Corresponding author (dpackwood@icems.kyoto-u.ac.jp)

**Abstract**

Advances in algorithms and hardware have enabled computers to design new materials atom-by-atom. However, in order for these computer-generated materials to truly address problems of societal importance, such as clean energy generation, it is not enough for them to have superior physical properties. It is also important for them to be adopted by as many users as possible. In this paper, we present a simple binary choice model for comparing catalyst materials on the basis of consumer preferences. This model considers a population of utility maximisers who select one of two materials by comparing catalytic turnover rates with sales prices. Through a mixture of numerical simulation and analytic theorems, we characterise the predictions of the model in a variety of regimes of consumer behavior. We also show how the model can be used as a guide for crafting policies for lowering catalyst prices in order to improve their market shares. This work represents a first step towards understanding how material properties should be balanced against production costs and consumer demand when designing new materials, an intellectual advance which may facilitate the spread of green materials in society.

**1. Introduction**

In materials science, the term 'computational materials design' indicates the task of constructing new materials on a computer in such a way that a physical property of interest is optimised [1, 2]. Typically, one will generate thousands of new materials by making subtle changes in atom positions and atom types, calculating the physical properties of each one along the way using a computational technique called density functional theory (DFT). With the increasing availability of computational power, improvements in DFT codes, and the emergence of machine learning optimization techniques over the last few years, the materials design process can, in principle, be performed quite efficiently [3, 4]. Now that this field is emerging from its nascent state, it is important to extend its scope beyond physical properties alone. Here, we ask the following question: *how can we form an overall assessment of a material, accounting for its physical properties, production costs, and consumer demand, in a consistent manner?* In order to answer this question, we clearly need to adopt concepts from the social sciences, particularly economics.

This question is very important when sustainability issues such as clean energy generation are used to motivate materials science research. In order for a material to address issues which affect society at large, it needs to be adopted as widely as possible. And in a society based on free enterprise, this requires that the material be purchased by as many potential users ('consumers') of that material as possible. A consumer makes their decision to purchase by weighing the properties of the material against the asking price of the firm. It is generally in the interest of the firm to reduce its asking price as much as possible, however it cannot charge a price lower than the production costs of the material or else it will lose money with each sale. For this reason, higher production costs are passed on to consumers in the form of higher prices, reducing demand for the material and limiting the degree to which it can address issues of broad societal importance.

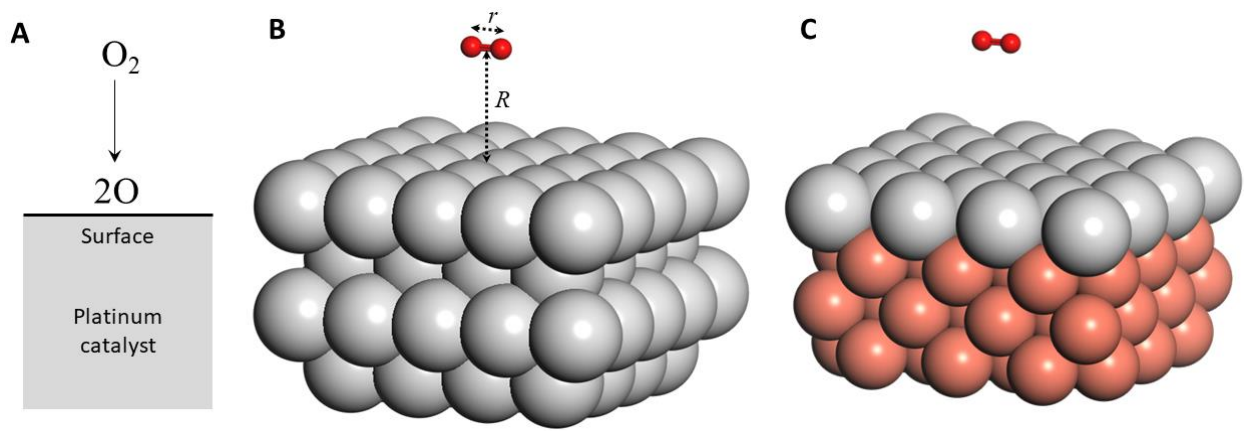

**Figure 1.** (A) Illustration of oxygen splitting by a platinum catalyst used in fuel cells. (B) Slab model a platinum catalyst (Pt catalyst) system used in this study. Grey and red spheres represent platinum and oxygen atoms, respectively. The red bar connecting the two oxygen atoms represents the chemical bond in the oxygen molecule. $R$ is the height of the oxygen molecule with respect to the catalyst surface and $r$ is the oxygen bond length. (C) Slab model of a copper-platinum catalyst (CuPt catalyst). Copper atoms are represented by brown-orange spheres. Slab models drawn with the Materials Studio Visualiser software [39].

The production cost issues associated with computationally designed materials can be understood by considering Figure 1, which shows two materials for catalyzing the oxygen splitting reaction ($O_2 \rightarrow 2O$, Figure 1A). This reaction is of critical importance in hydrogen fuel cells [5], where high reaction rates are required to ensure efficient energy generation. Figure 1B shows a pure platinum (Pt) catalyst, which is widely used in commercial hydrogen fuel cells at present [5]. Figure 1C shows another catalyst (CuPt) that was constructed by a computer. This catalyst consists of a platinum monolayer supported on a copper substrate. The efficiency of the catalyst is measured by a physical quantity called activation energy, which corresponds to the energy which must be supplied in order for the reaction to proceed. As will be described later, our DFT calculations predict an activation energy for the oxygen splitting reaction of 0.46 eV for Pt and 0.41 eV for CuPt. CuPt is therefore predicted to have marginally better catalytic properties than Pt. If CuPt were generated as part of a computational materials design protocol, it would be ranked ahead of Pt in the output.

For any type of material, there are two major components of production cost: raw material costs and opportunity costs. For the case of CuPt and Pt catalysts, raw material costs refer to the prices of copper and platinum metal. In order for a firm to fabricate CuPt, it would have to purchase these metals as a basic production input. The prices of metals fluctuate wildly in response to natural and geopolitical events, neither of which are easy to predict. However, the price of platinum metal is generally very high compared to copper - typically about $30 USD per gram for platinum compared

to less than 1 US cent per gram for copper [6]. For this reason, raw material costs for CuPt would be quite small compared to Pt catalyst due to the smaller amount of platinum metal required for production.

The second factor – opportunity costs – refer to the value of the opportunity lost by an engineer when they accept an employment contract by the firm to fabricate CuPt catalyst. In order for the firm to entice the engineer to accept the job contract, the firm must agree to pay a sum at least equal to this opportunity cost. Opportunity costs are therefore associated with (and often approximated by) wages [7]. Unlike raw material costs, the opportunity cost for fabricating a CuPt catalyst is expected to be quite large compared to the case of Pt. To see this, consider the (idealised) case of crystalline Pt and CuPt catalysts. For both cases, the preparation of the catalyst would start from the same step: the polishing of the platinum or copper substrate to obtain a single crystal. However, while the fabrication of Pt catalyst would finish at this polishing step, another step would be required for CuPt fabrication: deposition of a platinum monolayer onto the copper surface. This step requires skills beyond substrate polishing, including electron beam evaporation and electrochemical deposition, as well as electron diffraction and ellipsometry skills for evaluating monolayer quality. Such skills require advanced, graduate-level training, which are generally in short supply. The short supply of such skills, as well as the advanced training required to receive them, would mean that an engineer qualified for CuPt preparation could demand a much higher wage than one only qualified to polish platinum metal. In other words, such a skilled engineer would have to sacrifice other valuable opportunities in order to commit to fabricating CuPt. As a result, the reduced material costs for CuPt would be offset by the higher opportunity costs of production.

In situations where the material could serve as a public good, a case can be made for its production to be partly subsidized by government or other third parties in order to bring it to the market at a reasonable price. It is here where our question becomes highly relevant. At present, we lack any theoretical framework which can meaningfully evaluate a material through simultaneous account of material properties and opportunity costs. Without such a framework, it is difficult to estimate the size of the subsidy required for production of a material predicted by a computational design protocol, especially in cases when raw material costs and opportunity costs are high. Without such a framework, the ability for computational materials design to address societal issues such as sustainability will remain limited.

In this paper, we take a first step towards addressing our question by presenting a new model for performing comparative evaluations of catalyst materials. This model is built using concepts and techniques from microeconomics, and simultaneously accounts for catalytic turnover rates, production costs, and consumer demands in a seamless way. In short, the model involves setting up a competition between two fictitious firms which produce different materials. These firms compete for shares in a fictitious market, whose consumers consist of utility maximisers whose utility functions include both catalytic turnover rates and catalyst prices. In this work, we consider the case where the two firms produce Pt and CuPt catalyst, respectively. By a mixture of numerical simulation using time-series models for real metal prices, as well as analytic theorems for simplified cases, we characterize the market share behavior in a variety of regimes of consumer behavior and identify cases where CuPt is preferred by consumers. Moreover, by exploring market shares as a function of opportunity cost differences, we demonstrate how the model can be used to assist the development of policy for subsidizing material production. By considering the 'demand' side question of whether consumers will adopt the material, this work supplements other efforts which focus on modeling material production processes and optimising costs [8, 9]. Aside from a similar 'demand' side approach briefly mentioned in reference [10], we are unaware of similar works to ours in the materials science literature. This model could be applied to other types of materials produced by a materials design pipeline, allowing for one to broadly assess computer-generated materials beyond their physical properties alone.

Binary choice models are fundamental to microeconomics, but do not appear to have been applied to problems in materials science. As far as physical science is concerned, models involving discrete choices and utility-maximising agents can mainly be found in the social physics literature, particularly in works which aim to understand how collective social behavior arises [11]. General instances of these models have been studied by Durlauf [12], Radosz *et al.* [13], and Ostasiewicz *et al* [14]. A binary choice model applied to fashion choices has also been presented by Nakayama and Nakamura [15]. Binary and discrete choice models mediated by Ising-like graph interactions have been examined by Lowe *et al.* [16], Opoku *et al* [17], and Lee and Lucas [18]. Our work differs from these in that the emergent properties of the model (catalyst market share) are mediated by market forces rather than between agents on a microscopic level. On the other hand, our work can be seen as a first attempt to apply such models to solve problems in another area of physics, namely materials science.

This paper is organized as follows. Section 2 begins by describing our calculations of catalyst rate constants (section 2.1), followed by our binary choice model (2.2) and time-series models for describing material costs (2.3). Section 3 shows typical simulated results from our model under four representative regimes of consumer behavior (3.1), the effect of opportunity costs on market shares (3.2), and explores how the model can be used for policy purposes (3.3). Section 4 discusses the model shortcomings and future directions.

## 2. Methods

### *2.1. Catalyst materials and rate constant calculations*

We consider pure platinum (Pt) and copper-supported platinum monolayer (CuPt), as shown in Figure 1. Pt is widely used for catalyzing the $O_2$ dissociation reaction ($O_2 \rightarrow 2O$) in hydrogen fuel cells, as shown in Figure 1A. CuPt is a computer-generated material that we wish to compare to Pt as a potential catalyst.

CuPt was generated by creating a three-layer Cu slab terminated at the 001 face and placing Pt atoms on top of the hollow positions. The Pt catalyst structure was generated using a four-layer Pt slab terminated at the 001 face. We only consider the 001 faces only, as this is known to be the preferred face for $O_2$ dissociation for platinum [19]. The slab dimensions were 10.84 x 15.35 Å for CuPt and 11.10 x 11.10 Å for Pt. The slabs were relaxed using density functional theory (DFT) with the bottom two atom layers kept frozen. Relaxation was performed in VASP version 5.4.4 [20], using the PBE exchange-correlation function [20], PAW-PBE pseudopotentials, a 450 eV basis set cut-off, and 1 x 1 x 1 Γ-centered *k*-points grids.

Turnover rates for the $O_2$ dissociation reaction were computed using the Arrhenius equation:

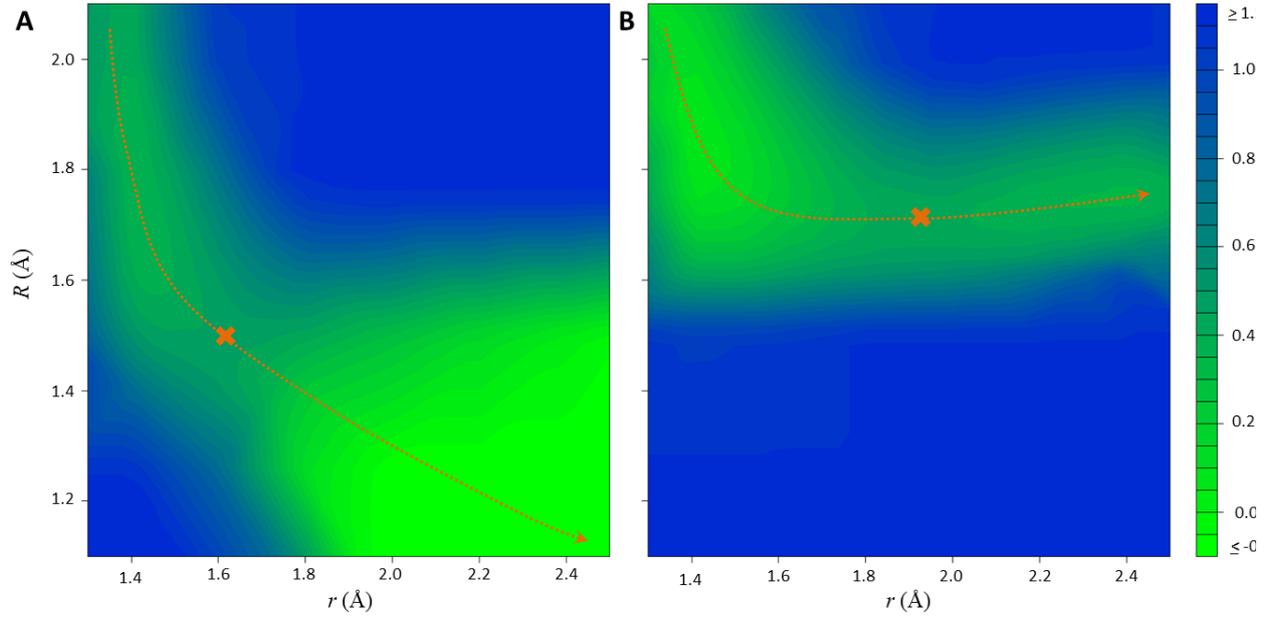

**Figure 2.** Interaction energies ($E_{\text{int}}(r, R)$) for (A) an oxygen molecule and a Pt catalyst surface and (B) an oxygen molecule and a CuPt catalyst surface. The dotted orange line represents the trajectory of the oxygen molecule during the oxygen splitting reaction. The orange x marks the location of the activation barrier. Plots drawn using R and the akima package [34, 40]

$$W_i = A_i \exp\left(-\frac{\varepsilon_i}{k_B T}\right), \qquad (1)$$

where $i$ denotes either CuPt or Pt, $k_B$ is the Boltzmann constant, $T$ is the temperature, $A_i$ is the frequency factor, and $\varepsilon_i$ is the activation energy, respectively. The frequency factors $A_i$ are assumed to be independent of catalyst type, which is acceptable because both catalysts should involve similar reaction geometries. We did not attempt to calculate the frequency factor explicitly, as it cancels in the model presented in the next section.

To obtain the activation energy, an $O_2$ molecule with a bond length $r$ was placed at various heights $R$ above the CuPt or Pt slab, in such a way that the O-O bond was parallel to the surface and intersected midway through one Pt-Pt bond (see Figure 1B). For given values of $r$ and $R$, the energy of interaction between the $O_2$ molecule and the catalyst slab was calculated according to the formula

$$E_{\text{int}}(r, R) = E(r, R) - E_{\text{ref}}, \qquad (2)$$

where $E(r, R)$ is the energy of the slab + molecule system as obtained from a single-point DFT calculation. $E_{\text{ref}}$ is a reference energy, which corresponds to the energy of the system when the catalyst and oxygen molecule are isolated from each other and not interacting. $E_{\text{ref}}$ can be obtained as $E_{\text{ref}} = E_{\text{slab}} + E_{\text{Ox}}$, where $E_{\text{slab}}$ is the energy of the slab in isolation and $E_{\text{Ox}}$ is the energy of an isolated $O_2$ molecule with its equilibrium bond length $r = 1.2$ Å. These energies were computed for various bond lengths $r$ between 1.2 Å and 2.5 Å, and various heights $R$ between 1.1 Å and 2.1 Å using DFT as implemented in VASP, with spin polarization, the rev-vdW-DF2 exchange-correlation functional [22], a 450 eV basis set cut-off, and 3 x 3 x 2 Γ-centered *k*-points grids. Note that equation (2) assumes that the relaxation rate of the surface is much slower than the $O_2$ dissociation rate.

A contour plot of the interaction energy is shown in Figure 2. From these plots, the activation energy $\varepsilon_i$ can be obtained by identifying the energy maximum along the lowest energy pathway between intact $O_2$ ($r = 1.2$ Å) and dissociated $O_2$ ($r = 2.1$ Å). In Figure 2, the locations of the activation barriers are indicated by the orange crosses. The activation energies are calculated to be $\varepsilon_{Pt} = 0.46$ eV for the Pt catalyst and $\varepsilon_{CuPt} = 0.41$ eV for the CuPt catalyst. At room temperature (300 K), this corresponds to a turnover rate ratio of $W_{CuPt}/W_{Pt} = 6.91$, which indicates a significant although not massive improvement of catalytic performance for CuPt over Pt. Similar results were reported in reference [23], which considered a monolayer platinum catalyst on an iron substrate.

*2.2. Model for comparative analysis of catalyst materials*

Our model for comparing Pt and CuPt catalysts runs as follows. Let $C_1$ and $C_2$ denote two firms which exclusively produce Pt and CuPt catalyst, respectively. $C_1$ and $C_2$ compete for shares in a market of *N* consumers, each of whom purchase exactly one unit of catalyst per unit time. Let $p_1$ and $p_2$ denote the prices at which $C_1$ and $C_2$ sell their respective catalysts. Moreover, let $m_1$ and $m_2$ denote the cost for $C_1$ and $C_2$ to produce one unit of their respective catalysts. $m_1$ and $m_2$ will contain both material costs and opportunity costs associated with production.

In this model, the competition between $C_1$ and $C_2$ for market shares proceeds in two steps. In the first step, each consumer declares which of the two catalysts they will purchase for given values of $p_1$ and $p_2$. In the second step, $C_1$ and $C_2$ adjust their prices in order to maximize the number of consumers who will purchase from them.

In order to deduce the outcome of the first step, we need to specify utility functions for each consumer. In our model, the gain in utility for consumer *k* when selecting firm $C_i$ ($i = 1, 2$) is

$$u_{ki} = \beta_k g(W_i) - \alpha_k p_i, \tag{3}$$

where *g* is an increasing, concave-down function, $W_i$ is the turnover rate of the catalyst produced by firm $C_i$, and $\beta_k$ and $\alpha_k$ are positive coefficients. $\beta_k$ is unitless and $\alpha_k$ has units of 1/price. The first term in equation (1) implies that consumers are only concerned with catalyst performance, and are indifferent towards other features such as material composition. Other implications of equation (3) are discussed in section 4. Consumer *k* will purchase a catalyst from firm $C_2$ if $u_{k2} \geq u_{k1}$, and will purchase from firm $C_1$ otherwise. The number of consumers who will purchase from firm $C_2$, given that catalyst prices are set at $p_1$ and $p_2$, is therefore

$$\begin{aligned} Q_2(p_1, p_2) &= \sum_{k=1}^{N} \mathbf{1}\big(\beta_k g(W_2) - \alpha_k p_2 > \beta_k g(W_1) - \alpha_k p_1\big) \\ &= \sum_{k=1}^{N} \mathbf{1}(\beta_k/\alpha_k > B), \end{aligned} \tag{4}$$

where $B = B(p_1, p_2)$ is defined as

$$B(p_1, p_2) = \frac{p_2 - p_1}{g(W_2) - g(W_1)}, \qquad (5)$$

and $\mathbf{1}(\beta_k/\alpha_k > B) = 1$ if $\beta_k/\alpha_k > B$ and is zero otherwise. The number of consumers who will purchase from firm $C_1$ for these prices is simply $Q_1(p_1, p_2) = N - Q_2(p_1, p_2)$. In order to transform equation (4) into a form more useful for calculations, we assume that $\beta_1, \beta_2, \ldots, \beta_N$ and $\alpha_1, \alpha_2, \ldots, \alpha_N$ are sequences of independent and identically distributed random variables. Let $P$ denote probability and observe that $P(\beta_k/\alpha_k > B)$ is independent of $k$. By the law of large numbers, we therefore have

$$Q_2(p_1, p_2)/N \to P(\beta_k/\alpha_k > B) \qquad (6)$$

with probability 1 as $N$ goes to infinity. Defining $q_2(p_1, p_2) = \lim_{N \to \infty} Q_2(p_1, p_2)/N$, and using the identity $P(\beta_k/\alpha_k > B) = 1 - \phi(B)$, where $\phi$ is the cumulative distribution of $\beta_k/\alpha_k$, we obtain

$$q_2(p_1, p_2) = 1 - \phi(B) \qquad (7)$$

and $q_1(p_1, p_2) = \phi(B)$. At this stage we say nothing about the function $\phi(B)$, aside from the fact that it must increase smoothly from 0 to 1 as $B$ changes from minus to plus infinity.

We now turn to the second step, in which $C_1$ and $C_2$ adjust their prices to maximize their market shares. This price adjustment must occur under the profit constraints

$$p_i q_i(p_1, p_2) - m_i q_i(p_1, p_2) \geq 0 \qquad (8)$$

where $i = 1, 2$, otherwise the firms would be operating at a loss. In order words, $p_i \geq m_i$ for both firms. From equation (7), it is clear that firm $C_2$ would like to make $\phi(B)$ as small as possible, which requires pushing $B$ towards negative values. In view of equation (5) and the profit constraint above, this involves $C_2$ setting $p_2$ to its smallest value, $m_2$. Conversely, $C_1$ would like to make $\phi(B)$ as large as possible, and this also requires setting $p_1$ to its smallest value, $m_1$. In the language of game theory, this price setting corresponds to a so-called Nash equilibrium strategy, in which neither firm has incentive to deviate from these prices as doing so would harm its market share. We therefore arrive at our final expressions for the market shares of the two firms:

$$q_2^* = 1 - \phi(B^*), \qquad (9)$$

where $B^* = (m_2 - m_1)/(g(W_2) - g(W_1))$, and $q_1^* = 1 - q_2^*$, where the asterisk denotes Nash equilibrium. Equation (9) is used for performing a comparative analysis of CuPt and Pt catalyst in the following sections.

*2.3. Production cost model*

In order to determine whether $C_2$ could gain a significant market share, the production cost differences $\Delta m = m_2 - m_1$ in equation (9) must be estimated. We proceed by splitting the marginal cost difference $\Delta m$ into two contributions:

$$\Delta m = \Delta m_r + \Delta m_s. \tag{10}$$

In equation (10), $\Delta m_r$ denotes material cost differences and $\Delta m_s$ denotes the difference in opportunity costs associated with CuPt fabrication compared to Pt fabrication. Of these two components, we first deal with $\Delta m_s$.

The opportunity cost of undergoing a particular activity is equal to the cost of the forgoing the best alternative to that activity. For the case of productive activities, opportunity costs can be understood in terms of the labour required for the production. The opportunity cost for an engineer to fabricate a particular catalyst is equal to the value that their labour would fetch on the labour market when applied to the next best job offer. Wage is often taken as a proxy for opportunity cost, and so in this work we describe $\Delta m_s$ as a wage difference. Namely, $\Delta m_s$ is the difference in the wage required to entice an engineer to fabricate CuPt catalyst compared to the wage required to entice them to fabricate ordinary Pt catalyst. In general, we expect $\Delta m_s$ to be positive and potentially quite large, due to the difficult operation of depositing a perfect monolayer of platinum atoms onto a copper surface, as well as subsequent quality checks through spectroscopy or electron diffraction measurements.

Material cost differences can be expressed with the formula

$$\Delta m_r = \left(M_{Cu} P_{Cu} + v M_{Cu} P_{Pt}\right) - M_{Pt} P_{Pt}. \tag{11}$$

The term in the brackets is the material cost of the CuPt catalyst. $M_{Cu}$ and $vM_{Cu}$ are respectively the mass of copper and mass of platinum used to produce the catalyst. In this formulation, the platinum mass is expressed as a fraction $v$ of the copper mass. $P_{Cu}$ and $P_{Pt}$ are the prices of copper and platinum, respectively, per unit mass. The second term is the material cost of the Pt catalyst, where $M_{Pt}$ is the mass of platinum used in a platinum fuel cell. We expect $\Delta m_r$ to be negative in general, because the material cost of producing a Pt catalyst would typically be higher than that of CuPt due to the larger amount of expensive material (platinum) involved.

In this work, $\Delta m_r$ is simulated using time-series models for iron and platinum prices. These models were fit to real data for the prices of copper and platinum during the period January 1 2016 – July 1 2022, as plotted in Figure 3 (source: [6]). Our fitting scheme is presented in the Appendix. Our scheme yields a GARCH(1,1) + ARMA(1,1) model for the (log-)price of platinum, which includes a non-stationary noise term as well as a deterministic trend. For the case of copper, we were unable to fit a simple time series model to the entire date range. We instead divided the time range into a 'pre-coronavirus' ($\leq$ 2020/3/22) and a 'coronavirus' ($\geq$ 2020/3/23) regime and fit the time series models separately for each regime. Application of our scheme yielded an ARIMA(0,1,0) model for the pre-coronavirus regime, and a GARCH(1,1) + ARMA(1,1) model for the coronavirus regime. In simulations of the copper price, the initial value for the coronavirus regime was set equal to the final value of the pre-coronavirus regime. These models for the platinum and copper prices yield time series which resemble the real data upon simulation, making them effective simulators for material prices

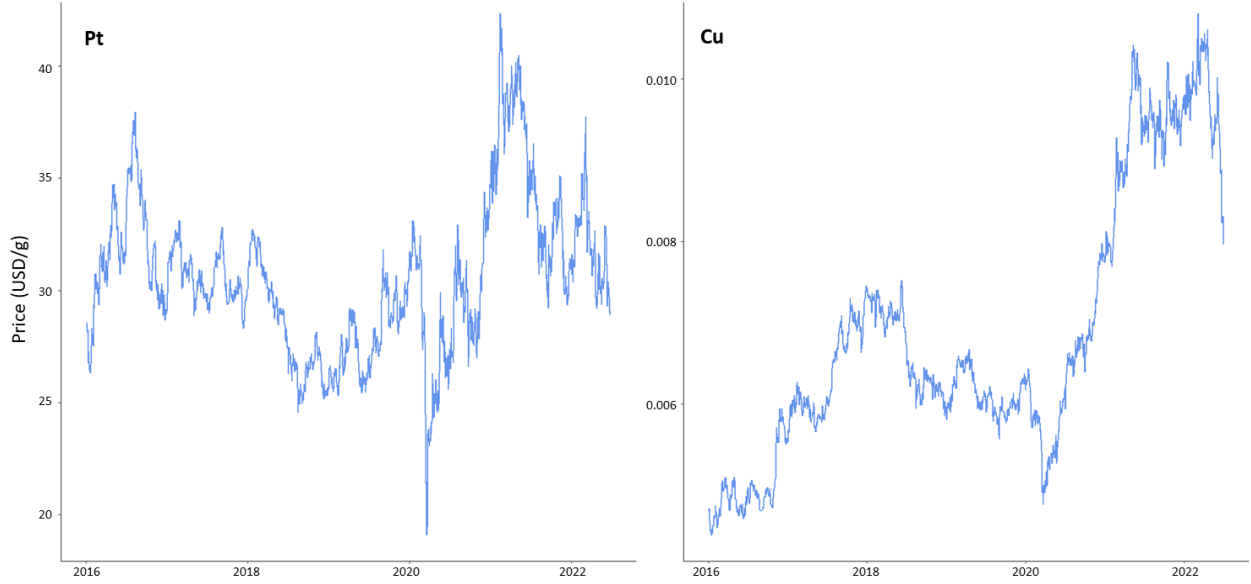

**Figure 3.** Time series data for platinum (Pt) and copper (Cu) prices. Data obtained from [22]. Plots drawn with R and the ggplot2 package [34, 41]

## 3. Results

Equilibrium market shares for firm $C_2$ can be simulated using equation (9) and the time-series models for the metal prices described above. However, in order to do this, it is first necessary to specify various parameter values. In the following, we set $g(W_i) = \ln W_i$, which implies diminishing utility gains with increasing catalytic performance. Logarithmic utility gains from product characteristics are commonly assumed for modeling demand for consumer products (e.g., [24] for the case of automobiles). We also assume that $\alpha_k \sim N(\mu_a, \sigma_a^2)$ and $\beta_k \sim N(\mu_b, \sigma_b^2)$, where $N$ denotes the normal distribution. This is a reasonable assumption given that many probability distributions for positive random variables can be approximated by normal distributions in certain limits. Importantly, this assumption allows us to make the normal approximation for the distribution of $\beta_k/\alpha_k$ [25]. Namely, $\beta_k/\alpha_k \sim N(\mu, \sigma^2)$, where,

$$\mu = \mu_b / \mu_a \tag{12}$$

and

$$\sigma^2 = \frac{\sigma_b^2}{\mu_a^2} + \frac{\mu_b^2 \sigma_a^2}{\mu_a^4} \tag{13}$$

$\mu$ and $\sigma$ both have units of price. In equation (11), we set $M_{Pt} = 50$ g, which is close to the mass of platinum used in real platinum fuel cells at present [5], and $M_{Cu} = M_{Pt}\rho_{Cu}/\rho_{Pt}$, where $\rho_{Cu}$ and $\rho_{Pt}$ are respectively the densities of copper and platinum. $v$ is set to 0.1, which is suitable for thin-film catalysts.

|  | HCHP | LCHP | HCLP | LCLP |
|---|---|---|---|---|
| $\mu_a$ (1/USD) | 5 | 5 | 2.5 | 2.5 |
| $\mu_b$ | 1000 | 500 | 1000 | 500 |
| $\sigma_a$ (1/USD) | 2.5 | 2.5 | 1.25 | 1.25 |
| $\sigma_b$ | 500 | 250 | 500 | 250 |
| $\Delta m_s$ (USD) | 2000 | 2000 | 2000 | 2000 |

**Table 1** Parameters used for the four regimes of consumer behavior.

We first illustrate the framework by simulating (9) under four regimes of consumer behavior:

- High catalyst performance concerns, high price concerns (HC-HP)
- Low catalyst performance concerns, high price concerns (LC-HP)
- High catalyst performance concerns, low price concerns (HC-LP)
- Low catalyst performance concerns, low price concerns (LC-LP)

The specific parameter settings for these regimes are shown in Table 1. These regimes differ in the means and variances of the coefficients $\beta_k$ and $\alpha_k$. Under the two HP regimes (HC-HP, LC-HP), consumers tend have a strong aversion to spending, making the mean of $\alpha_k$ large. Under the two HC regimes, (HC-HP, HC-LP), consumers strongly value catalyst performance, making the mean of $\beta_k$ large. These parameter regimes are selected to illustrate the range of outcomes possible with our framework. The question of whether or not they are realistic will be considered in section 4.

*3.1. Simulated market share behavior*

Market shares for firm $C_2$ under the four regimes are shown in Figure 4 for a typical metal price trajectory and a fixed opportunity cost difference to $2500 per catalyst. We first consider the two HP regimes (HC-HP and LC-HP). While the CuPt catalyst is the marginally better catalyst in terms of catalytic rate constants, the high opportunity costs associated with its production must be paid by consumers, who are averse to spending. Thus, even in the HC-HP regime, the market share of firm $C_2$ is typically quite small. The market share of $C_2$ shrinks significantly on passing from the HC-HP to LC-HP regime, and is essentially zero during periods when the Pt price is low. In this situation, consumers do not place high value on catalyst performance, making it irrational for them to purchase from firm $C_2$ unless platinum prices are particularly high. These observations show that, despite having the better catalytic properties, the ability for CuPt to address societal issues will be poor in unfavorable market conditions.

We now consider the two LP regimes (HC-LP and LC-LP). As expected, the market share for firm $C_2$ becomes quite large in the HC-LP regime, as consumers are less averse to shouldering the larger

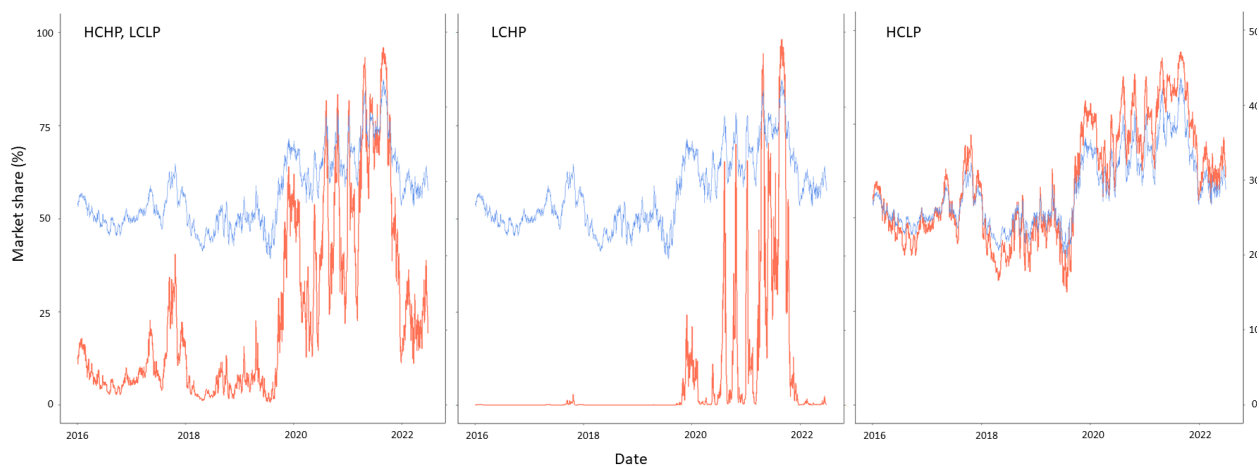

**Figure 4**. Market shares for the CuPt catalyst under the four regimes of consumer behavior (red), for a fixed realization of platinum prices (blue). Plots drawn with R and the ggplot2 package [34,41]

production costs in order to obtain the marginally more effective catalyst, CuPt. Interestingly, the market share in the HC-LP regime closely follows the price of platinum, meaning that $C_1$'s tenuous market share rises and shrinks in inverse proportion to the price of platinum. Under these regimes, CuPt could achieve a wide enough user base for addressing issues of societal importance.

While the above results are not surprising, a non-trivial observation can be made from Figure 4. Namely, that the LC-LP regime yields identical results to the HC-HP regime. Mathematically, this occurs because the parameters of the LC-LP regime are precisely half those of the HC-HP regime, and this factor of a half cancels when calculating the parameters of the cumulative distribution function in (4). This observation has a non-trivial implication: despite outward differences in behavior, consumers with a high concerns towards performance and price will make identical choices as those with low concerns towards performance and price, at least on a macroscopic level. Low price aversion in the market is therefore not a sufficient condition for CuPt to achieve widespread user adoption.

*3.2. Effect of opportunity cost*

The proceeding discussion assumed a fixed opportunity cost difference $\Delta m_s$ and fixed realization of Cu and Pt prices. However, a more valuable application of this model is to predict the expected market share for $C_2$ as a function of $\Delta m_s$. This application will allow us to develop policies for expanding the market share of CuPt catalyst, even under unfavorable market conditions.

Figure 5 plots the expected market share under each of the four regimes described above. The expected market share was obtained as the time-average of 100 independent simulations of equation (9). In all four cases the market share decays as a sigmoid-shaped curve. In particular, the market shares stays at around 100 % until $\Delta m_s$ reaches about $1100 - $1200 USD per unit and quickly decay afterwards. The rate of decay is fastest for the LC-HP region, in which the spending-averse consumers see little benefit in purchasing the CuPt catalyst and are repelled by the prospect of shouldering $C_2$'s growing production costs. For this regime, the market share of $C_2$ is essentially zero for opportunity cost differences of $4000 USD or more. For opposite reasons, the rate of decay is slowest for the HC-LP regime: these consumers value catalyst performance and are less averse to

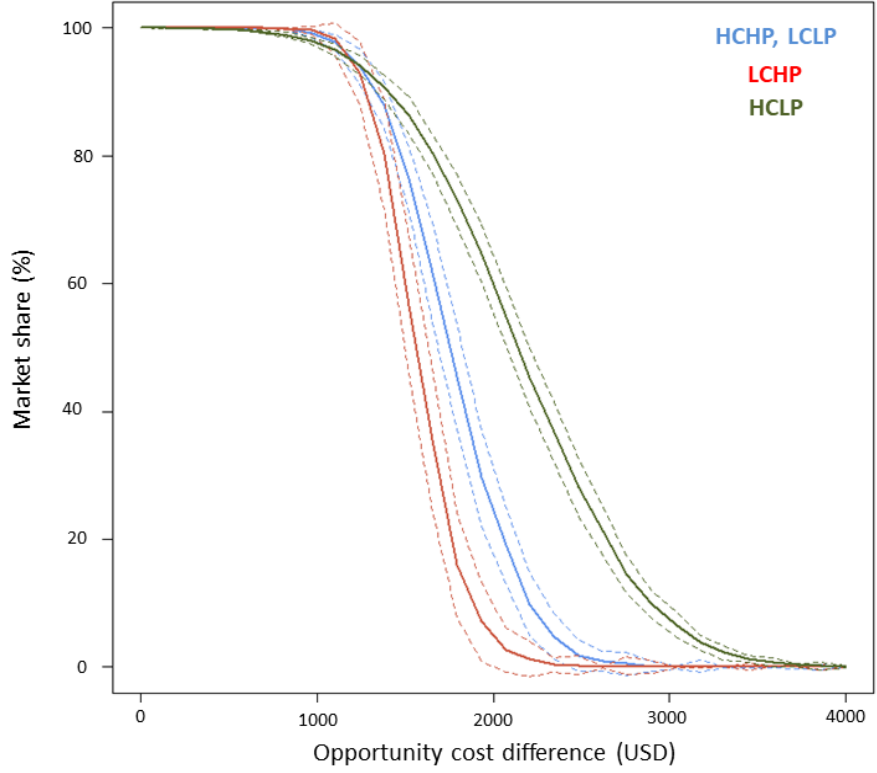

**Figure 5.** Expected market shares for CuPt catalyst as a function of opportunity cost differences for CuPt versus Pt fabrication. Dotted curves correspond to standard deviation about the mean. Different colours correspond to different regimes of consumer behavior. Plot drawn with R [34].

spending, making them more tolerant to shouldering higher production costs of $C_2$. The HC-HP and LC-LP represent an intermediate case, and both produce identical results due to the reasons described above.

In order to interpret the plot in Figure 5, we turn Theorems 1 and 2 presented in Appendix 2. These theorems provide analytic results for the simplified case where

$$B^* = \frac{\Delta m_s + \Delta m_r}{\ln W_2/W_1} \tag{14}$$

is treated as a normally distributed random variable with mean

$$a = \frac{\Delta m_s + \langle \Delta m_r \rangle}{\ln W_2/W_1}, \tag{15}$$

where $\langle \Delta m_r \rangle$ is a time-averaged material cost difference, $a > 0$, and a finite standard deviation. This is a simplification because, in general, it will not be possible to represent $B^*$ in this way when realistic time-series models for $\Delta m_r$ are used. Theorem 1 shows that for small enough $a$ and large enough $\mu$, the market share for firm $C_2$ is approximately 1, i.e., $\phi(B^*) \sim 0$ and $q_2^* = 1 - \phi(B^*) \sim 1$ for almost all realisations of $B^*$. We can use this result to explain the observation from Figure 5 that $C_2$'s market share stays near 100 % until $\Delta m_s$ is around \$1100 - \$1200 USD. Let us approximate equation (11) as $\Delta m_r \approx -M_{Pt}P_{Pt}$, the negative of the material costs of a single unit of Pt catalyst. Thus, for *a* to be close to zero, we must have $\Delta m_s \approx -\langle \Delta m_r \rangle$, or $\Delta m_s \approx M_{Pt}\langle P_{Pt} \rangle$. By averaging the time-series data in Figure 3 over time, we can estimate $\langle P_{Pt} \rangle$ = $30.5 USD/g, which yields $M_{Pt}\langle P_{Pt} \rangle$ = $ 1525 USD for $M_{Pt}$ = 50 g. Thus, according to the simple model, market shares will be near 100 % for $\Delta m_s \approx$ $1525 USD. This value corresponds well to the observation of around $\Delta m_s \approx$ $1100 - $1200 USD considering the approximations made the analysis, judtifying the use of Theorem 1 to interpret these results. We can therefore conclude that opportunity cost differences in the order of material price differences are required for $C_2$ to obtain a near-100 % market share.

Theorem 2 states that, for the case of the simplified model, the expected market share of firm $C_2$ is 50 % when $a = \mu$, or

$$\Delta m_s^{max} = \frac{\mu_b}{\mu_a} \ln \frac{W_2}{W_1} - \langle \Delta m_r \rangle, \tag{16}$$

Equation (16) can be interpreted as the maximum opportunity cost difference allowed for firm $C_2$ to be competitive with $C_1$ (in the sense of acquiring at least 50 % of the market share). Setting $\langle P_{Pt} \rangle$ = $30.5 USD/g once again, we obtain $\Delta m_s^{max}$ = $1912, $1719, $2299, and $1912 USD/g for the HC-HP, LC-HP, HC-LP, and LC-LP regimes, respectively. This correspond to the values $\Delta m_s^{max}$ of $1755, $1565, $2150, and $1755 USD/g respectively, obtained from the curves in Figure 5. This correspondence is acceptable considering the simplifications of this analysis, supporting the use of equation (16) to understand the results in Figure 5. $\Delta m_s^{max}$ increases as catalyst performance ($W_2$) and consumer preferences for catalyst performance ($\mu_b$), and decreases at consumer price concerns ($\mu_a$) increase. Approximating $\langle \Delta m_r \rangle \approx -M_{Pt}\langle P_{Pt} \rangle$ shows that $\Delta m_s^{max}$ increases linearly with the average price of platinum.

*3.3 Policy development*

As was shown in section 3.1, CuPt is not necessarily going to penetrate deeply into the market despite being a better catalyst that Pt. This would be problematic from the point-of-view of national energy and carbon emissions goals, which require that novel materials be adopted by as widely as possible in society.

In our model, the only way for firm 2 to boost its market share is to decrease its marginal costs of producing CuPt catalyst. One way it could do this is to improve the efficiency of the CuPt fabrication process. This would have two effects. The first effect would be to reduce material costs, as less of the expensive material could be used. However, this effect would be minor for the case of CuPt, where the amount of platinum used is miniscule. It may also reduce electricity costs or other running costs, although these were not considered in the current model. The second effect would be to reduce opportunity costs. The value of the opportunity lost due to fabricating CuPt would become less as the fabrication process becomes simpler. However, this does not mean that the firm would reap an immediate benefit from improvement efficiencies. Indeed, wages tend to be 'sticky', meaning that the labour market is slow to adjust to shocks such as technological innovations [26]. Thus, if the firm were to shock the labor market by suddenly simplifying its fabrication procedure, it would in the short term continue to pay the same opportunity costs as before.

For the case of a CuPt-producing firm, there is therefore little it can do by itself to reduce production costs in the short-term. However, the results in the previous section, particularly the interpretations provided by Theorems 1 and 2, suggests ways in which the effective production costs of the firm could be reduced with the help of a third party such as government or other agents such as investors. Indeed, let us decompose opportunity cost differences as

$$\Delta m_s = \Delta m_s^* + \delta. \tag{17}$$

In equation (17), $\Delta m_s^*$ represents a 'desired' opportunity cost difference for firm 2 to reach a target market share, and $\delta$ represents the remainder. For example, if firm 2 desires a near 100 % market share, it could set $\Delta m_s^* = M_{Pt}\langle P_{Pt}\rangle$, which is equal to around $1525 USD/catalyst in the example considered here. Then, if a third party were to pay the remainder $\delta$, production costs would effectively become $\Delta m = \Delta m_r + \Delta m_s^*$, allowing the firm to achieve the desired market share. On the other hand, if firm 2 desires at least a 50 % market share, then it could set $\Delta m_s^* = \Delta m_s^{max}$, as defined in equation (16). For the cases considered in this paper, this would correspond to $\Delta m_s^*$ in the range of $1565 - $2150 USD/catalyst. The remainder $\delta$ would again be covered by other sources.

The model presented here could therefore be used as a guide for policy makers and investors. They could determine market conditions from real market surveys, and from data on the labor market they could estimate the true opportunity cost for scientists and engineers to produce CuPt catalyst or other functional materials in question. By entering into negotiation with industry, they could then decide upon a value $\Delta m_s^*$ which the firms would have to bear, and the remainder that they would have to pay, either through tax revenue or other sources.

## 4. Discussion

Is the model presented here 'correct', in any sense? Computational materials scientists will often judge a calculation by simply comparing it to experimental data. However, our model cannot be compared to real data, as there has never been an instance in history where CuPt and Pt have undergone duopolistic competition in a catalyst market. We should therefore assess the model by considering the validity of its assumptions. The major assumptions to consider are the choice of the utility function in (3), the use of independent random coefficients to model the population of consumers, and the idea that firms compete on price to maximize market share.

Utility functions similar to the one in equation (3) are often used for modeling consumer demand, particularly ones where product characteristics enter logarithmically and prices enter linearly. One of the most influential studies of this type was published by Berry *et al* in 1995, who succeeded to predict market shares for automobiles on the basis of such utility functions [24]. Other studies which discuss log-linear product characteristics include [27] and [28]. At this stage, there is no reason to suspect that such assumptions (logarithmic dependence on characteristics, linear dependence on prices) do not hold well for catalyst markets either. A more serious shortcoming of our utility function is the absence of an 'unobserved' component, that is, some property of the catalyst which is known to consumers and producers, but not known by the modeler. We have ignored this here because the unobserved component needs to be correlated with price, which seriously complicates subsequent analysis. Our utility function also ignores budget constraints of the consumers, which is important for providing consumers with the option of not purchasing any catalyst at all. The utility function in (3) should therefore be considered as a minimum model for consumer behavior, from which more adequate utility functions could be developed in the future.

In our model, the consumers differ according to the values of the coefficients in the utility function. In order to model this variation, we treated these coefficients as independent random variables with a known probability distribution. This assumption is convenient and commonly used throughout economics (e.g., [29] and [30]). However, in realistic situations we need to consider the possibility that consumers can influence each other's decisions, which bring into question the assumption of independence. We should also consider the joint distribution between the coefficients for price and catalytic turnover rates, as these coefficients could be correlated as well. In a realistic application of our model, the joint probability distribution for these coefficients could be estimated from market surveys. It is obviously beyond the scope of the present study to conduct such a survey, however for future work this might be considered.

The final assumption to discuss relates to the way in which the two firms compete. In our model, we considered a two-step process, in which consumers state which catalyst they would prefer to purchase given prices in the first step, and then firms respond by adjusting their prices in a non-cooperative manner to maximize sales in the second step. The majority of oligopoly models suppose that firms either adjust prices or quantities in order to maximize profit (these are known as the Bertrand scenario and the Cournot scenario, respectively [31]), however there is no consensus in the economics literature as to which of these is more valid. Unfortunately, the Bertrand scenario is analytically non-tractable in our case, because price enters the profit function in a complicated, non-linear way. Moreover, the Cournot scenario is difficult with our two-step process, which obliges the firms to fulfil specific consumer demands for given prices. However, a large number of authors have argued that firms act to maximize sales rather than profits, as managers are often incentivized to do so by company owners [32]. Our model would therefore fall within this literature.

In light of the above discussion, the assumptions used in our model are meaningful and conform to acceptable microeconomic theory. However, there is one other aspect of our analysis which should be discussed, namely the decomposition of production costs into raw material costs and opportunity costs. With many materials scientists and engineers focusing on modeling material production process, one might wonder how the material production process itself enters into this decomposition. Actually, this is implicit. On the one hand, the amount of material used by a production process is implicit in the material costs. Material costs could be expanded to include electricity cost and other utility prices, although here we have not done this. On the other hand, the complexity of the production process is bound up in opportunity cost, as the more complex a process is the more valuable the opportunity lost by undertaking the process becomes. However, for opportunity cost to be a useful concept one must be able to quantify in some way. Usually this is done by considering the wages required to entice an appropriately qualified person to undertake the job, although it is unclear whether such economic intuition applies to the labour market for engineers and scientists. In any case, in addition to incorporating models of material production processes directly into a computational materials design protocol, it might also be fruitful to estimate opportunity costs by characterizing the labour market for scientists and engineers. This would clearly be a topic for future research.

A final point to consider is whether our discussion of opportunity costs is complete. In addition to opportunity costs associated with labour, a firm would also need to consider depreciation of capital. This concept is made more concrete in the area of Keynesian economics under the banners of 'user costs' and 'supplementary costs' (see [33] for an accessible discussion). The user cost of a production process is the opportunity cost associated with using production equipment in the present instead of delaying its use for the future. User cost can be high if the equipment used for production deteriorates rapidly with use, and if the manager expects to obtain a higher return if they were to shift production to a future date. User cost might be particularly important in the present context, as the equipment used for depositing and checking monolayer deposition (such as electron beam evaporators and electron diffraction equipment) have short lifetimes and are not designed for heavy use. Because such equipment deteriorates so rapidly with use, we would expect 'supplementary costs', which refer to the expected degradation of equipment irrespective of use, to be low compared to user costs. The opportunity costs discussed in previous sections should therefore contain a user cost component, in addition to wage. The relative contribution of wage and user costs in the overall opportunity costs associated with materials production should be taken up in future research.

## 5. Conclusion

In order for materials science to truly address problems of societal importance such as sustainability, it needs to embrace methods from social science in addition to well-established tools such as density functional theory and machine learning. In this paper, we have presented a microeconomic model for comparing catalyst materials as a first step towards this goal. This model incorporates catalytic turnover rates, production costs, and consumer demand parameters in a consistent way, and allows one to explore how two catalyst materials will perform relative to each other in a fictitious marketplace. Through a mixture of numerical simulations and analytic theorems, we showed that near-100 % market shares for a catalyst can be obtained when opportunity cost differences between catalyst production are equal and opposite to material cost differences. We also deduced an analytic condition for the catalyst to obtain at least a 50 % market share and hence be competitive. From these results, it appears possible to construct policy for investment or subsidies to improve the market share of the catalyst material and hence improve its ability to address problems of societal importance. This model could easily be applied to other types of materials as well, and with further improvements may facilitate the realization of a green materials revolution.

**Acknowledgements**

This work was supported by K. Matsushita Foundation grant 21-G09, the Sompo Environment Foundation, Japan Society for the Promotion of Science Kakenhi grants 18K14126, 19H04574, and 21K05003, and internal funding from the Graduate School of Advanced Integrated Studies in Human Survivability and the Institute for Integrated Cell-Material Sciences at Kyoto University.

**Author contributions**

KS: conceptualization (equal), data curation (equal), formal analysis (equal), investigation (equal), methodology (equal), funding acquisition (equal), validation (equal), visualization (equal), writing – original draft (supporting), writing – review and editing (supporting).

DP: conceptualization (equal), data curation (equal), formal analysis (equal), investigation (equal), methodology (equal), project administration (main), resources (main), funding acquisition (equal), validation (equal), visualization (equal), supervision (main), writing – original draft (main), writing – review and editing (main).

**Competing Interests**

The authors declare no competing interests.

**Data Availability Statement**

The copper and platinum price data can be downloaded from reference [6]. Data generated from our models is available from the authors upon reasonable request.

**Appendix 1. Time-series model details**

Time-series models for the platinum and copper price data (Figure 3) were build in a step-wise manner according to the scheme in Figure 6. In the following, we let $y_t$ denote the price of the metal (platinum or copper) at time $t$ and $z_t = \ln y_t$. All hypothesis tests and model parameter estimations were performed within the R statistical environment [34].

*Platinum time-series model*

*Step i*. The KPSS test was used to determine whether $z_t$ possessed a unit root. For the case of $z_t$, we were able to reject the null hypothesis that there is no unit root (test statistic = 2.32), suggesting that $z_t$ is not stationary. The KPSS test was performed with the function "ur.kpss" in the package "tseries" [35].

*Step ii*. We attempted to fit an ARIMA model to the time series $z_t$. The model which minimized the AIC (Akaike Information Criterion) was ARIMA(0,1,0), which is equivalent to a random walk with Gaussian-distributed steps. In order to confirm whether this is an appropriate model for $z_t$, a Jarque-Bera test was conducted. The null hypothesis (that $z_t - z_{t-1}$ is normally distributed) was rejected (*p*-value < 2.2 x $10^{-16}$). Based on this result, we reject the ARIMA(0,1,0) model as a model for $z_t$. Finally, we conducted the Breusch-Pagan test to check for heteroskedasticity. The null hypothesis (homoskedasticity in the residuals of $z_t$ with respect to the ARIMA(0,1,0) model above) was rejected (*p*-value = 9.5 x $10^{-4}$), suggesting heteroskedasticity in the time-series $z_t$. The ARIMA model was fit using the function "arima" with the package "stats". The Jarque-Bera test was performed using the function "jarque.bera.test" in the package "tseries". The Breusch-Pagan test was performed using the function "bptest" in the package "lmtest" [36].

*Step iii*. The presence of heteroskedasticity suggests that we attempt to fit a GARCH(*p*, *q*) model to $z_t$. The GARCH(*p*, *q*) model has the form $r_t = \mu + \varepsilon_t$, where $E(\varepsilon_t^2 \mid F_{t-1}) = \sigma_t^2$,

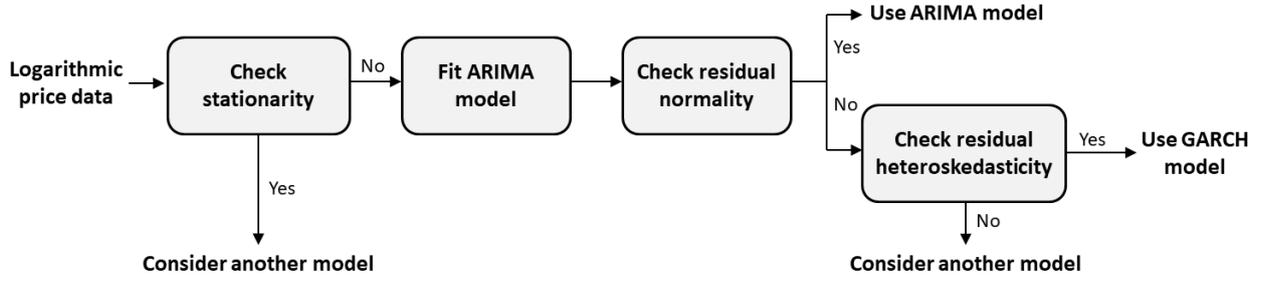

**Figure 6.** Flow diagram for selecting time-series model

$$\sigma_t^2 = \omega + \sum_{i=1}^{p} \alpha_i \varepsilon_{i-1}^2 + \sum_{i=1}^{q} \beta_i \sigma_{i-1}^2 , \tag{18}$$

and $F_{t-1}$ denotes the sigma field generated by $\varepsilon_j$, $j = 0, \ldots, t - 1$. A GARCH(1,1) model fitted to $z_t$ resulted in $\mu = 6.85$ (*p*-value = 0), $\omega = 1.33 \times 10^{-4}$ (*p*-value = $1.0 \times 10^{-6}$), $\alpha_1 = 0.89$ (*p*-value = 0), and $\beta_1 = 0.10$ (*p*-value = $8.17 \times 10^{-3}$). This model achieved an AIC of -2.77. Finally, a GARCH(1,1) + ARMA(1,1) model was fit to $z_t$, resulting in $\mu = 6.79$ (*p*-value = 0), $\omega = 1.0 \times 10^{-6}$ (*p*-value = 0.34), $\alpha_1 = 3.5 \times 10^{-2}$ (*p*-value = 0), $\beta_1 = 0.96$ (*p*-value = 0), an autoregression parameter of 0.99 (*p*-value = 0) and a moving average parameter of $1.77 \times 10^{-2}$ (*p*-value = 0.46). This model achieved an AIC of -5.75. While the GARCH(1,1) + ARMA(1,1) model contained two statistically insignificant parameters, we opted to use it over the simpler GARCH(1,1) model because its simulated sample paths appeared to resemble the original platinum price data ($y_t$) when exponentiated and plotted. These models were fit using the functions "ugarchspec" and "ugarchfit" in the package "rugarch" [37].

*Copper time-series model*

For the copper prince data, we created a composite model by fitting two independent time-series models for dates prior and post 22 March 2020, respectively. We have two reasons for doing this. Firstly, the residuals of an ARIMA model fit to the logarithmic copper price ($z_t$) across the entire data rate had non-normal and autocorrelated residuals, suggesting that this ARIMA model inappropriate. On the other hand, when applied to the entire date range, the Breusch-Pagan test did not detect any heteroskedasticity, suggesting that a GARCH model would not fit well. Thus, our scheme in Figure 6 tells us to consider another model. Secondly, it is well understood that copper demand increased dramatically following the outbreak of the coronavirus pandemic in March 2020. The market forces acting on copper prices before and after March 2020 will therefore be qualitatively different, and therefore should be treated with different models.

*Step i ($\leq$ 22 March 2020data).* The KPSS test was used to test for stationarity. For the logarithmic copper price ($z_t$), the null hypothesis that there is no unit root could be rejected (test statistic = 6.55). This suggests that $z_t$ is not stationary.

*Step ii ($\leq$ 22 March 2020 data).* An ARIMA model was fit to the time series $z_t$. We obtained an ARIMA(0,1,0) model as the one which minimises the AIC. This results suggests that $z_t$ can be modeled as a random walk with Gaussian-distributed steps. Furthermore, the Jarque-Bera test applied to the residuals of $z_t$ ($z_t - z_{t-1}$) yielded a *p*-value of 0.15, suggesting that the null hypothesis should not be rejected and supporting the random walk as a model for $z_t$ in this regime.

*Step i (> 22 March 2020 data).* The KPSS test showed that the null hypothesis that $z_t$ has no unit root could be rejected (test statistic = 6.70), suggesting that $z_t$ is not stationary.

*Step ii (> 22 March 2020 data).* An ARIMA model was fit to $z_t$. We obtained an ARIMA(1,1,1) model was the one which minimised the AIC. However, the Jaque-Bera test suggested that the residuals of this model were non-normal (*p*-value = 3.7 x $10^{-9}$), suggesting that this is an inappropriate model for $z_t$ in this regime.

*Step iii (> 22 March 2020 data).* The Breusch-Pagan test to check for heteroskedasticity of $z_t$. The null hypothesis (homoskedasticity in the residuals of $z_t$ with respect to the ARIMA(1,1,1) model above) was rejected (*p*-value = 2.2 x $10^{-5}$), suggesting heteroskedasticity in the time-series $z_t$.

*Step iv (> 22 March 2020 data).* Due to the presence of heteroskedascity above, we decided to fit a GARCH(*p*, *q*) model to $z_t$. An ARMA(1,1) + GARCH(1,1) fitted to $z_t$ resulted in a model with $\mu$ = 0.78 (*p*-value < 2.2 x $10^{-16}$), $\omega$ = 6.5 x $10^{-5}$ (*p*-value = 0.07), $\alpha_1$ = 8.92 x $10^{-2}$ (*p*-value = 0.06), $\beta_1$ = 0.60 (*p*-value = 1.48 x $10^{-3}$), an autoregression parameter of 1.00 (*p*-value < 2.2 x $10^{-16}$) and a moving average parameter of -0.10 (*p*-value = 0.11).

Copper prices were therefore modeled as an ARIMA(0,1,0) process in the first regime ($\leq$ 22 March 2020) and an ARMA(1,1) + GARCH(1,1) process in the second regime (> 22 March 2020). In simulations of the composite model, the initial condition of the ARMA(1,1) + GARCH(1,1) process was set equal to the value of the ARIMA(0,1,0) process on 22 March 2020.

**Appendix 2. Analytic results for simplified model**

Consider the random variable

$$Q_n = 1 - \phi_n(Z_n), \tag{19}$$

where $Z_n \sim N(a_n, b_n^2)$, $a_n > 0$, and $\phi_n : \mathbf{R} \to [0, 1]$ is the cumulative normal distribution function

$$\phi_n(x) = \frac{1}{2}\left[1 + \text{erf}\left(\frac{x - \mu_n}{\sigma_n \sqrt{2}}\right)\right], \tag{20}$$

where $\mu_n$ and $\sigma_n$ represent the mean and standard deviation parameters of $\phi_n$.

The conceptual correspondence between the model in the main paper (equation (9)) and equation (20) is as follows. $a_n$ corresponds to $(\Delta m_s + \langle \Delta m_r \rangle)/(g(W_2) - g(W_1))$ where $\Delta m_s$ represents opportunity cost differences and $\langle \Delta m_r \rangle$ represents time-averaged material cost differences. The random variation about $a_n$ represents stochastic fluctuations of the material cost differences about $\langle \Delta m_r \rangle$.

**Theorem 1.** Consider a sequence of random variables $Q_1$, $Q_2$, ..., as defined in (19), such that $a_n$ is decreasing, $a_n \to 0$, $\mu_n$ is strictly increasing, and $\mu_n \to \infty$. Furthermore, suppose that $b_n$ and $\sigma_n$ are both non-increasing. Then for any $\delta > 0$ and every $\varepsilon > 0$ there is an $n_0$ such that for all $n > n_0$, the event

$$E_n = \{\phi_n(Z_n) > \delta\} \qquad (21)$$

occurs with probability less than $\varepsilon$.

*Proof.* This result follows from the Borel-Cantelli lemma, providing that

$$S = \sum_{n=1}^{\infty} P(E_n) \qquad (22)$$

converges, where $P(E_n)$ is the probability of event $E_n$. To prove the convergence of $S$, observe that

$$P(E_n) = P(Z_n > \phi_n^{-1}(\delta)) = 1 - F_n(\phi_n^{-1}(\delta)), \qquad (23)$$

where $\phi_n^{-1} : [0,1] \to \mathbf{R}$ is the inverse map of $\phi_n$ and $F_n : \mathbf{R} \to [0,1]$ is the cumulative distribution of $Z_n$. The first equality holds because $\phi_n(x)$ is strictly increasing in $x$. To simplify notation, let $y_n = \phi_n^{-1}(\delta)$. Under the assumptions of the theorem, $y_n$ strictly increases as $n$ increases. Moreover, let

$$1 - F_n(y_n) = \frac{1}{2}\operatorname{erfc}(z_n) \qquad (24)$$

where

$$z_n = \frac{y_n - a_n}{b_n^2 \sqrt{2}}. \qquad (25)$$

Under the assumptions of the theorem, $z_n$ also increases strictly as $n$ increases. Thus, for large $n$, we can apply the approximation (see [38])

$$\operatorname{erfc}(z_n) = \frac{1}{\sqrt{\pi}} e^{-z_n^2} \frac{1}{z_n} + O(z_n^{-1}). \qquad (26)$$

We now apply the ratio test to confirm the convergence of (22). From (24) and (28) we obtain

$$\lim_{n \to \infty} \frac{1 - F_{n+1}(z_{n+1})}{1 - F_n(z_n)} = e^{-(z_{n+1}^2 - z_n^2)} \frac{z_n}{z_{n+1}} < 1, \qquad (27)$$

which completes the proof.

**Theorem 2.** Suppose that for some $n$, $a_n = \mu_n$. Then $E(Q_n) = \frac{1}{2}$.

*Proof.* We drop the subscript $n$ for simplicity. By the symmetry of the cumulative normal distribution, we have the identity

$$\phi(z) + \phi(2\mu - z) = 1. \tag{28}$$

for all $z$. Conditioning on the random variable $I_\mu = \mathbf{1}(Z > \mu)$, where $\mathbf{1}$ is the indicator function, and observing that $P(I = 1) = P(I = 0) = \frac{1}{2}$, we obtain

$$\begin{aligned} E(\phi(Z)) &= E\left(E\left(\phi(Z) | I_\mu\right)\right) \\ &= \frac{1}{2} \int_\mu^\infty \phi(z) f_\mu(z) dz + \frac{1}{2} \int_\mu^\infty \phi(2\mu - z) f_\mu(z) dz \end{aligned} \tag{29}$$

where $f_\mu$ is the probability density of $Z$ conditioned in $Z > \mu$. The second equality follows from the assumption that $a = \mu$ and that the probability density of $Z$ is symmetric about $\mu$. Inserting equation (28) into (29) yields the result.